\documentclass[10pt,aps,prl,twocolumn,superscriptaddress,floatfix,showpacs,longbibliography]{revtex4-1}

\usepackage{graphicx}
\usepackage{amsfonts}
\usepackage{amssymb}
\usepackage{amsmath}
\usepackage{txfonts}
\usepackage{lipsum}
\usepackage{color}
\usepackage{wasysym}
\usepackage{hyperref}
\usepackage{bbold}
\usepackage{etoolbox} 
\usepackage[normalem]{ulem}
\usepackage{mathtools} 
\usepackage{multirow} 
\usepackage{hhline}
\usepackage{mathrsfs} 
\usepackage{soul}

\newcommand{\av}[1]{\ensuremath{\left\langle #1 \right\rangle}}

\definecolor{green}{rgb}{0,0.6,0.1}

\usepackage{letltxmacro}
\LetLtxMacro{\oldsqrt}{\sqrt}
\renewcommand{\sqrt}[2][\mkern8mu]{\mkern-6mu\mathop{}\oldsqrt[#1]{#2}}

\usepackage{url}
\definecolor{indigo(dye)}{rgb}{0.0, 0.25, 0.42}
\usepackage{placeins}

\begin{document}

\title{
Orbital Isotropy of Magnetic Fluctuations in Correlated Electron Materials \\
Induced by Hund's Exchange Coupling
}

\author{Evgeny A. Stepanov}
\affiliation{I. Institute of Theoretical Physics, University of Hamburg, Jungiusstrasse 9, 20355 Hamburg, Germany}
\affiliation{Theoretical Physics and Applied Mathematics Department, Ural Federal University, Mira Str. 19, 620002 Ekaterinburg, Russia}

\author{Yusuke Nomura}
\affiliation{RIKEN Center for Emergent Matter Science, 2-1 Hirosawa, Wako, Saitama 351-0198, Japan}

\author{Alexander I. Lichtenstein}
\affiliation{I. Institute of Theoretical Physics, University of Hamburg, Jungiusstrasse 9, 20355 Hamburg, Germany}
\affiliation{Theoretical Physics and Applied Mathematics Department, Ural Federal University, Mira Str. 19, 620002 Ekaterinburg, Russia}

\author{Silke Biermann}
\affiliation{CPHT, CNRS, Ecole Polytechnique, Institut Polytechnique de Paris, F-91128 Palaiseau, France}
\affiliation{Coll\`ege de France, 11 place Marcelin Berthelot, 75005 Paris, France}

\begin{abstract}
Characterizing non-local magnetic fluctuations in materials with strong electronic Coulomb interactions remains one of the major outstanding challenges of modern condensed matter theory.
In this work we address the spatial symmetry and orbital structure of magnetic fluctuations in perovskite materials.
To this aim, we develop a consistent multi-orbital diagrammatic extension of dynamical mean field theory, which we apply to an anisotropic three-orbital model of cubic $t_{2g}$ symmetry.
We find that the form of spatial spin fluctuations is governed by the local Hund's coupling. 
For small values of the coupling, magnetic fluctuations are anisotropic in orbital space, which reflects the symmetry of the considered $t_{2g}$ model. 
Large Hund's coupling enhances collective spin excitations, which mixes orbital and spatial degrees of freedom, and magnetic fluctuations become orbitally isotropic.
Remarkably, this effect can be seen only in two-particle quantities; single-particle observables remain anisotropic for any value of the Hund's coupling. 
Importantly, we find that the orbital isotropy can be induced both, at half-filling and for the case of $4$ electrons per lattice site, where the magnetic instability is associated with different, antiferromagnetic and ferromagnetic modes, respectively. 
\end{abstract}

\maketitle

An accurate description of many-body effects in multi-orbital systems represents a challenging task for theoretical condensed matter physics. 
In addition to collective charge, spin and superconducting fluctuations that are present already in effective single-orbital systems, realistic materials possess orbital degrees of freedom, which greatly enhances the wealth of the physical phenomena displayed by such materials.
In strongly interacting electronic systems these fluctuating degrees of freedom become entangled, and it is difficult to foresee which collective effects govern the physics of the system.

Perovskite materials with partially filled $t_{2g}$ orbitals reveal a high degree of anisotropy and may serve as an attractive playground for studying the interplay between orbital and spin degrees of freedom~\cite{M1}.
Prominent examples are LaTiO$_3$, SrRuO$_3$ and, to a lesser degree, Sr$_2$RuO$_4$. 
In these materials, strong magnetic fluctuations have been revealed by inelastic neutron scattering experiments~\cite{PhysRevLett.85.3946, BUSHMELEVA2006491, PhysRevLett.123.017202,  PhysRevLett.83.3320, PhysRevLett.122.047004, PhysRevB.103.104511}.
For LaTiO$_3$, it has been argued that the joint effect of the superexchange interaction and the Hund's exchange coupling leads to a disordered orbital ground state~\cite{PhysRevLett.85.3950}.
Despite antiferromagnetic superexchange interactions, SrRuO$_3$ is an itinerant ferromagnet.
It undergoes a Curie transition at $T_{c}=160$~K~\cite{RevModPhys.84.253}, which is accompanied by a distortion of the ideal cubic perovskite structure (see e.g. Ref.~\onlinecite{PhysRevB.74.094104}).
While the formation of the ferromagnetic state in this material can be well captured within {\it ab initio} density functional calculations~\cite{doi:10.1063/1.361618, PhysRevB.53.4393, PhysRevB.56.2556}, 
description of itinerant ferromagnetism above $T_{c}$ is challenging and requires consideration of long-range collective electronic fluctuations.
In Sr$_2$RuO$_4$, coupling of Hund's exchange to orbital degrees of freedom was found to be responsible for the magnetic fluctuations~\cite{PhysRevLett.106.096401, doi:10.1146/annurev-conmatphys-020911-125045, PhysRevLett.112.127002, Boehnke_2018, PhysRevB.100.125120, Acharya2019}.
Quite generally, anisotropies in correlated systems may favor instabilities of various nature, such as orbital ordering~\cite{OrbitalOrder}, Peierls instabilities~\cite{CorrPeierls}, strong crystal-field splitting~\cite{d1crystal}, and Fermi-surface instabilities related to the Pomeranchuk effect~\cite{Pomeranchuk1}. 
The theoretical description of this phenomenology is hindered by the fact that local approximations to  electronic Coulomb correlations tend to overestimate anisotropies of a system. 
Indeed, taking into account long-range fluctuations may drastically change the physical picture~\cite{Pomeranchuk2}. 
Nevertheless, even most recent works on correlated multi-orbital systems are based on dynamical mean-field (DMFT) calculations~\cite{PhysRevX.10.041002, kang2020infinitelayer, PhysRevB.103.054503}, illustrating the lack of computationally tractable approaches beyond the local picture. 

In this work we address the problem of spin fluctuations in perovskite materials close to a magnetic instability, using a realistic three-orbital $t_{2g}$ model.
We find that non-local spin fluctuations enhanced by large Hund's exchange coupling strongly reduce the orbital anisotropy of the perovskite structure. 
As a consequence, magnetic fluctuations become isotropic in orbital space, as we show both at half-filling, as well as for the case of $4$ electrons per lattice site.
These results illustrate the important role that the local Hund's coupling plays not only for the local spin physics, but also for the symmetry and orbital structure of spatial magnetic excitations.
A second important result of our work is the design of a minimal consistent multi-orbital extension of the recently introduced dual triply irreducible local expansion ($\text{D-TRILEX}$) approach~\cite{PhysRevB.100.205115, PhysRevB.103.245123}.
The $\text{D-TRILEX}$ has a diagrammatic structure similar to $GW$~\cite{GW1, GW2, GW3}, which allows for a self-consistent consideration of the feedback of collective electronic fluctuations onto the single-particle quantities and {\it vice versa}.
However, in contrast to $GW$, our method allows for a simultaneous and unambiguous accounting for leading collective electronic effects including magnetic fluctuations in a simple partially bosonized way~\cite{PhysRevLett.121.037204, PhysRevB.99.115124, PhysRevB.100.205115}.
Moreover, these many-body effects can be considered as true long-range fluctuations without any spatial restrictions, which is a decisive advantage over cluster extensions of DMFT~\cite{PhysRevB.58.R7475, PhysRevB.62.R9283, PhysRevLett.87.186401, RevModPhys.77.1027, doi:10.1063/1.2199446, RevModPhys.78.865}.
Finally, the $\text{D-TRILEX}$ approach accounts for the exact local three-point vertex corrections at both sides of the $GW$-like diagrams for the self-energy and polarization operator, which preserves correct orbital structure of spatial fluctuations.
These vertices are crucial for describing the isotropic nature of the spin fluctuations described above: in their absence, this physics is not even qualitatively captured.

{\it Model ---} 
We start with a realistic $t_{2g}$ tight-binding model for the perovskite materials described by the three-orbital Hamiltonian
\begin{align}
{\cal H} = - \sum_{ij,l,\sigma} t^{ll}_{ij} c^{\dagger}_{il\sigma} c^{\phantom{\dagger}}_{jl\sigma} + \frac12 \sum_{i,ll'} \left( U^{\rm ch}_{ll'} n^{\phantom{\dagger}}_{il} n^{\phantom{\dagger}}_{il'} +  U^{\rm sp}_{ll'} m^{\phantom{\dagger}}_{il} m^{\phantom{\dagger}}_{il'} \right),
\label{eq:Hamiltonian}
\end{align}
where operator $c^{(\dagger)}_{il\sigma}$ annihilates (creates) an electron with spin projection ${\sigma=\{\uparrow, \downarrow\}}$ on site $i$ and orbital ${l=\{1, 2, 3\}}$. 
The anisotropy of this model originates from hopping parameters $t^{ll}_{ij}$ that are diagonal in the orbital space and have the following structure in momentum ({\bf k}) space~\cite{Pavarini_2005}
\begin{align}
t_{ll}({\bf k}) = \epsilon + 2t_\pi ({\cal C}_\alpha + {\cal C}_\beta)+2t_\delta {\cal C}_\gamma +4t_\sigma {\cal C}_\alpha {\cal C}_\beta,
\label{eq:dispersion}
\end{align}
where $\epsilon$ is the center of bands and ${{\cal C}_{\alpha} = \cos k_{\alpha}}$. 
For simplicity, we introduce three non-equivalent $\alpha$, $\beta$, $\gamma$ indices, where the first two are defined by the orbital label ${l=\{\alpha\beta\}}$ with ${1 = yz}$, ${2 = zx}$, and ${3 = xy}$. 
The last index $\gamma$ takes the remaining value among ${\{x, y, z\}}$. 
In this model, orbital degrees of freedom are tied to a spatial motion of the electrons, because the latter can hop only within the same orbital in a strictly defined direction, which is different for every considered orbital. The
$t_{\pi, \delta, \sigma}$ matrix elements describe the main hopping processes that provide the $t_{2g}$ symmetry. 
We choose ${t_\pi=1}$, which defines the energy scale of the system, and a realistic value for ${t_\delta=0.12}$ for the SrVO$_3$ perovskite~\cite{Pavarini_2005}. 
We note that $t_\sigma$ plays the role of $t^\prime$ in a two-dimensional model for cuprates and shifts the van-Hove singularity (vHS) away from the Fermi level. 
The presence of the vHS at the Fermi energy results in a peak in the density of states, which enhances correlation effects in the system.
Thus, for the half-filled case (${\langle N_i \rangle = 3}$ electrons per site) we preserve the particle-hole symmetry for $t_{2g}$ bands and set $t_\sigma=0$. 
For the case of ${\langle N_i \rangle = 4}$ we choose the positive value for ${t_\sigma=0.35}$~\cite{Pavarini_2005}, which ensures that the vHS again appears at the Fermi level~\cite{PhysRevB.86.085117}.

The on-site charge and spin density operators are defined as ${n_{il}=n_{il\uparrow}+n_{il\downarrow}}$ and ${m_{il}=n_{il\uparrow}-n_{il\downarrow}}$, where ${n_{il\sigma}=c^{\dagger}_{il\sigma}c^{\phantom{\dagger}}_{il\sigma}}$. 
The interaction is parametrized in the Kanamori form~\cite{10.1143/PTP.30.275} with intraorbital $U$ and interorbital $U'$ Coulomb interactions, and the Hund's coupling $J$. 
This parametrization is rotationally invariant provided ${U'=U-2J}$. Given that the matrix of hopping amplitudes is diagonal in orbital space, we consider only the density-density part of the Kanamori interaction  
\begin{align}
2U^{\rm ch} &= 
\begin{pmatrix}
U & U^{*} & U^{*} \\
U^{*} & U & U^{*} \\
U^{*} & U^{*} & U
\end{pmatrix}, ~~~~
2U^{\rm sp} = 
\begin{pmatrix}
-U & -J & -J \\
-J & -U & -J \\
-J & -J & -U
\end{pmatrix},
\end{align}
where ${U^{*} = 2U'-J}$. 
This expression for the interaction between charge and spin densities can be obtained rewriting the intraorbital Coulomb potential in the Ising-like form 
\begin{align}
U n_{il\uparrow}n_{il\downarrow} = \frac{U}{4} \Big(n_{il}n_{il} - m_{il}m_{il}\Big). 
\end{align}
As has been shown recently, this decoupling provides a relatively good result for the self-energy~\cite{PhysRevLett.119.166401, PhysRevB.100.205115}.

\begin{figure}[t!]
\includegraphics[width=0.9\linewidth]{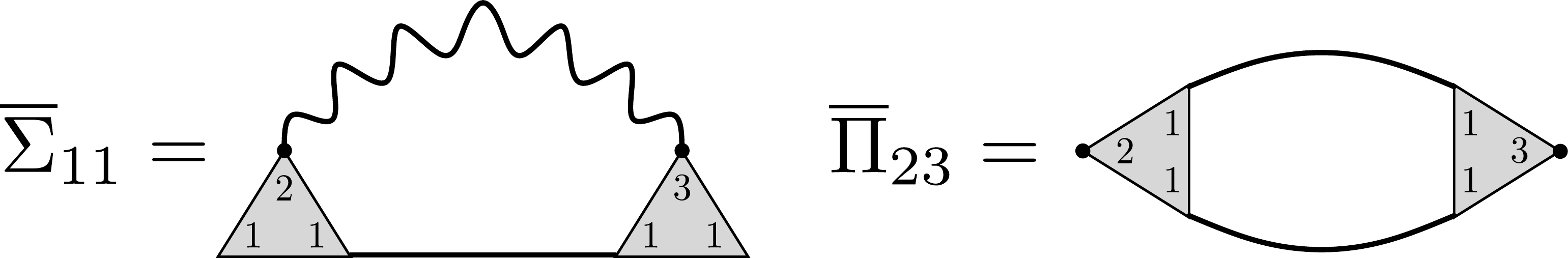}
\caption{\label{fig:Sigma_Pi} Diagrams for the non-local self-energy $\overline{\Sigma}_{ll}$ and polarization operator $\overline{\Pi}_{ll'}$. Grey triangles correspond to full local vertex functions $\Lambda_{ll'}$ of DMFT impurity problem. Wiggly line depicts the renormalized interaction $W_{ll'}$. Other bold lines are dressed non-local Green's functions $\overline{G}_{ll}$. Labels $l=\{1, 2, 3\}$ are orbital indices. Summation over internal orbital indices is implied.}
\end{figure}

{\it Many-body effects ---}
The non-interacting part of the problem~\eqref{eq:Hamiltonian} is highly anisotropic. 
We find, however, that many-body interactions can drastically change this property of the system. 
In this work collective electronic fluctuations are taken into account via the simplest consistent diagrammatic extension of DMFT~\cite{RevModPhys.68.13}, which yet allows to consider desirable lowest-order vertex corrections without heavy numerical efforts. 
This theory is formulated as a multi-orbital extension of the $\text{D-TRILEX}$ approach~\cite{PhysRevB.100.205115, PhysRevB.103.245123} that has been introduced recently as a simplified version of the dual boson (DB) theory~\cite{Rubtsov20121320, PhysRevB.90.235135, PhysRevB.93.045107, PhysRevB.94.205110, PhysRevB.100.165128, PhysRevB.102.195109}. 
Both methods use DMFT as a starting point for diagrammatic expansion. 
Thus, the local self-energy $\Sigma^{\rm imp}_{ll}(\nu)$ and polarization operator $\Pi^{\rm imp}_{l'l''}(\omega)$ are given by an effective impurity problem of DMFT-type. 
To avoid double-counting issues, the diagrammatic part of the theory that accounts for spatial correlation effects is formulated in a dual space. 
To this aim, we  perform a transformation of initial fermionic variables and exactly integrate out the local impurity problem~\cite{PhysRevB.100.205115}. This allows to construct an analog of the Almbladh functional~\cite{doi:10.1142/S0217979299000436} in the dual space ${\Psi[\overline{G}, W^{\varsigma}, \Lambda^{\hspace{-0.05cm}\varsigma}] = \frac12 \overline{G}^{\phantom{|}}_{ll} \Lambda^{\hspace{-0.05cm}\varsigma}_{ll'} W^{\varsigma}_{l'l''} \Lambda^{\hspace{-0.05cm}*\,\varsigma}_{l''l} \overline{G}^{\phantom{|}}_{ll}}$, which guarantees consistency between single- and two-particle quantities by means of the non-local self-energy and polarization operator
\begin{align}
&\overline{\Sigma}_{ll}(k) 
= - \sum_{q,l'l'',\varsigma} \overline{G}^{\phantom{1}}_{ll}(k+q) \,
\Lambda^{\hspace{-0.05cm}\varsigma}_{ll'}(\nu,\omega) \,  W^{\varsigma}_{l'l''}(q) \, \Lambda^{\hspace{-0.05cm}* \, \varsigma}_{l''l}(\nu,\omega)
\label{eq:Sigma} 
\\
&\overline{\Pi}^{\varsigma}_{l'l''}(q)  
= \,2\sum_{k,l} \Lambda^{\hspace{-0.05cm}* \, \varsigma}_{l'l}(\nu,\omega) \, \overline{G}^{\phantom{\varsigma}}_{ll}(k+q) \, \overline{G}^{\phantom{\varsigma}}_{ll}(k) \, \Lambda^{\hspace{-0.05cm}\varsigma}_{ll''}(\nu,\omega) \label{eq:Pi}
\end{align}
Labels ${k=\{{\bf k},\nu\}}$ and ${q=\{{\bf q},\omega\}}$ describe momentum ${\bf k}$ (${\bf q}$) and Matsubara fermion $\nu$ (boson $\omega$) frequency dependence. ${\overline{G}_{ll}(k) = G_{ll}(k) - g_{ll}(\nu)}$, where $g_{ll}(\nu)$ is the local part of the lattice Green's function $G_{ll}(k)$. $W^{\varsigma}_{l'l''}(q)$ is the renormalized interaction in the charge (${\varsigma={\rm ch}}$) and spin (${\varsigma={\rm sp}}$) channel. 
These quantities can be obtained self-consistently via standard Dyson equations
${G_{ll}^{-1}(k) = i\nu + \mu - t_{ll}({\bf k}) -  \Sigma^{\phantom{1}}_{ll}(k)}$ and ${W^{\varsigma\,-1}_{l'l''}(q) = U^{\varsigma\,-1}_{l'l''} - \Pi^{\varsigma}_{l'l''}(q)}$, where $\mu$ is the chemical potential, and ${\Sigma^{\phantom{1}}_{ll}(k) = \Sigma^{\rm imp}_{ll}(\nu) + \overline{\Sigma}^{\phantom{1}}_{ll}(k)}$ and ${\Pi^{\phantom{1}}_{l'l''}(q) = \Pi^{\rm imp}_{l'l''}(\omega) + \overline{\Pi}^{\phantom{1}}_{l'l''}(q)}$ are the total self-energy and polarization operator, respectively~\footnote{In order to make the first implementation of the multi-orbital $\text{D-TRILEX}$ theory simple, we do not consider the renormalization of the non-local self-energy by a ``dual'' denominator~\cite{PhysRevB.100.205115}, because it does not qualitatively affect described effects.}.
In this way, the $\text{D-TRILEX}$ theory provides an equal footing description of collective charge and spin fluctuations. 
The susceptibility $X^{\varsigma}_{ll'}(q)$ in the corresponding channel, which is an experimentally observable quantity, can also be obtained via Dyson's equation ${X^{\varsigma\,-1}_{ll'}(q) = \Pi^{\varsigma\,-1}_{ll'}(q) - U^{\varsigma}_{ll'}}$.

Finally, it is worth noting that the introduced improved $GW$-like form for the non-local self-energy~\eqref{eq:Sigma} and polarization operator~\eqref{eq:Pi} additionally accounts for vertex corrections at both sides of the corresponding diagrams~\footnote{For this reason, in Ref.~\onlinecite{PhysRevB.100.205115} the method was called TRILEX$^2$ approximation for the DB theory.}.  
$\Lambda^{\hspace{-0.05cm}\varsigma}_{ll'}(\nu,\omega)$ is the full local three-point vertex given by the DMFT impurity problem, and the quantity ${\Lambda^{\hspace{-0.05cm}*\,\varsigma}_{ll'}(\nu,\omega) = \Lambda^{\hspace{-0.05cm}\varsigma}_{l'l}(\nu+\omega,-\omega)}$ is introduced to simplify notations.
As Fig.~\ref{fig:Sigma_Pi} demonstrates, this form of the diagrams allows to preserve correct orbital symmetry of electronic fluctuations.
Indeed, the orbital structure of both lattice sites that are connected by the non-local Green's function $\overline{G}$ is considered in a symmetric way, which is missing in the original TRILEX approach~\cite{PhysRevB.92.115109, PhysRevB.93.235124}.   
It should be noted that the full local vertex function $\Lambda_{ll'}$ serves as the bare interaction vertex in the renormalized perturbation $\text{D-TRILEX}$ theory~\cite{PhysRevB.100.205115, PhysRevB.103.245123}. 
Therefore, the introduced diagrammatic structures~\eqref{eq:Sigma} and~\eqref{eq:Pi} do not contradict the exact Hedin form for the self-energy and polarization operator~\cite{GW1}. 
As has been clarified in Ref.~\onlinecite{PhysRevB.94.205110}, both of these expressions can be identically rewritten in the conventional Hedin form that contains a non-local vertex function at one side of the diagram.

A particular symmetry of the considered model~\eqref{eq:Hamiltonian} allows us to use a simplified version of the multi-orbital $\text{D-TRILEX}$ approach~\cite{ToBePublished}, where the vertex function $\Lambda_{ll'}$ and renormalized interaction $W_{l'l''}$ are taken in the density-density form and depend on two orbital indices instead of four. 
This makes the dressed Green's function $G_{ll}$ diagonal in the orbital space and thus anisotropic. 
However, as we shall see later, the initial single-particle anisotropy of the model~\eqref{eq:dispersion} does not necessarily extend to two-particle quantities. 
Indeed, although the Green's function is diagonal, the presence of vertex corrections $\Lambda_{ll'}$ leads to non-diagonal contributions to the non-local polarization function~\eqref{eq:Pi}. 
Further, a matrix structure of the Dyson equation for the renormalized interaction $W_{ll'}$ and the susceptibility $X_{ll'}$ even more entangles orbital and spatial degrees of freedom. 
In this way, strong non-local collective fluctuations, which are magnetic in our particular case, can destroy the spatial anisotropy in the orbital space.
This observation suggests to reconsider the commonly believed mean-field-based statement that correlations usually tend to increase the anisotropy of a system. 

{\it Orbital isotropy of magnetic fluctuations ---}
Remarkably, we find that the strength and orbital structure of spatial magnetic fluctuations are controlled by the value of the local Hund's coupling $J$.
To illustrate this point, let us first consider the interacting three-orbital model~\eqref{eq:Hamiltonian} at half-filling with ${U=4}$ and temperature ${T=1/2}$. 
For the specified parameters, the leading eigenvalue (l.e.) $\lambda$ of the Dyson equation for the  susceptibility $X_{ll'}$ indicates that strongest collective excitations in the system correspond to a magnetic instability channel.
We observe that for a relatively small ${J=0.2}$, the l.e. of magnetic fluctuations is not very large (${\lambda=0.78}$). 
In this case, the diagonal (intraorbital) part of the spin susceptibility $X^{\rm sp}_{ll}$ presented in Fig.~\ref{fig:W_hf}\,$a$ for the $yz$ orbital is much larger than the non-diagonal (interorbital) one shown in the Supplemental Material (SM)~\cite{SM}.
Moreover, the ${X^{\rm sp}_{11}(q_{x}, q_{y}; q_{z}=0, \omega=0)}$ component of the susceptibility is highly anisotropic in momentum space and is almost dispersionless along $q_{x}$ direction. This spatial structure of the spin susceptibility originates from the orbital symmetry of $t_{2g}$ hopping processes~\eqref{eq:dispersion}. 
The same symmetry also leads to the identical $q_{y}$ and $q_{z}$ momentum dependence of $X^{\rm sp}_{11}({\bf q})$. 
Importantly, all three diagonal components 
$X^{\rm sp}_{11}$, $X^{\rm sp}_{22}$, and $X^{\rm sp}_{33}$ of the susceptibility show a similar behavior in momentum space with a pronounced dispersionless structure along $q_x$, $q_y$ and $q_z$ directions, respectively. This result indicates that for small Hund's coupling, orbital degrees of freedom are anisotropic.

\begin{figure}[t!]
\includegraphics[width=0.95\linewidth]{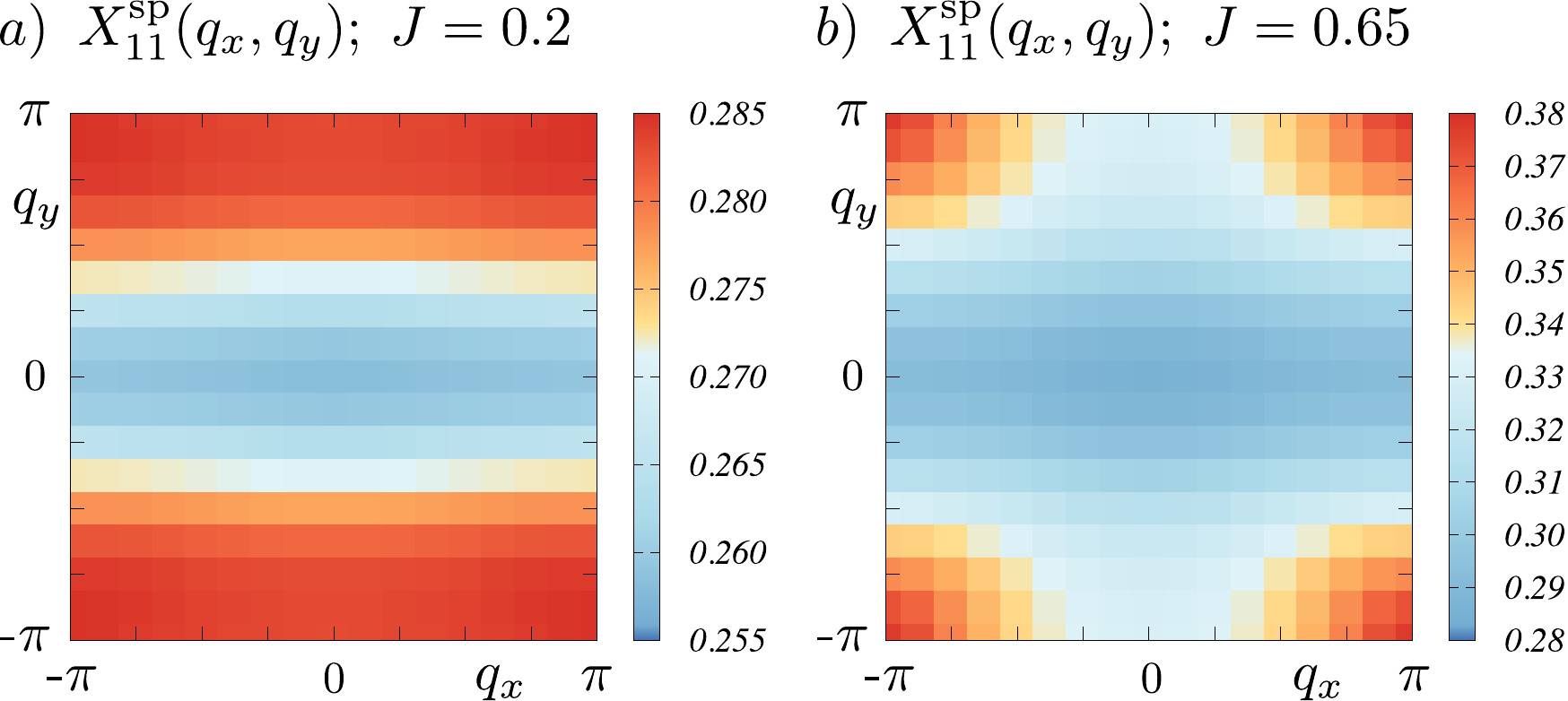}
\caption{\label{fig:W_hf} The absolute value of the diagonal $yz$ orbital component of the spin susceptibility $X^{\rm sp}_{11}(q_{x}, q_{y}; q_{z}=0, \omega=0)$ obtained for the half-filled $t_{2g}$ model for $U=4$. Color bars show the strength of $X^{\rm sp}$.
(a) In the case of small Hund's coupling $J=0.2$, the diagonal component of the susceptibility is highly anisotropic and is almost dispersionless along $q_{x}$ direction. (b) Increasing the Hund's coupling to $J=0.65$, intraorbital spin fluctuations become isotropic with a pronounced antiferromagnetic behavior depicted by the largest value of $X^{\rm sp}_{11}$ at corners of the Brillouin zone. 
}
\end{figure}

Increasing the value of the Hund's coupling to ${J=0.65}$, the magnetic l.e. approaches unity (${\lambda=0.99}$), which indicates that spin fluctuations are strongly enhanced~\footnote{
In order to get such large leading eigenvalue we reduce the threshold for the self-consistency to 10 iterations. Otherwise, it would require a more precise adjustment of the temperature and the Hund's coupling.}. 
This can also be concluded from Fig.~\ref{fig:W_hf} comparing the amplitude of the susceptibility for two considered cases of $J$.
Moreover, at this large value of the Hund's coupling interorbital components of $X^{\rm sp}$ (see SM~\cite{SM}) become comparable to intraorbital ones (Fig.~\ref{fig:W_hf}\,$b$). 
This is the first signature of the isotropic orbital behavior of magnetic fluctuations. 
A proximity of the l.e. to unity indicates that all orders of an effective perturbation expansion given by the Dyson equation contribute almost equally to the total $X^{\rm sp}$.
This leads to a more thorough mixing of orbital and spatial degrees of freedom in the susceptibility.
As shows Fig.~\ref{fig:W_hf}\,$b$, this results in a highly isotropic form of spin fluctuations with a clearly distinguishable antiferromagnetic (AFM) behavior. 
Interorbital components of the susceptibility remain isotropic in momentum space~\cite{SM}.
This means, that orbital degrees of freedom are no more tied to a specific spatial direction defined by hopping parameters~\eqref{eq:dispersion} of the considered model. As a consequence, collective fluctuations in the magnetic channel become orbitally isotropic.

Remarkably, the non-local part of the dressed Green's function shown in the SM~\cite{SM} remains spatially anisotropic for both considered cases of the Hund's coupling. This can be explained by the fact that the bare Green's function is highly anisotropic and isolates only the anisotropic contribution from the non-local self-energy~\eqref{eq:Sigma}, despite that the renormalized interaction $W^{\rm sp}_{l'l''}$ can be isotropic. Therefore, the orbital isotropy induced by strong magnetic fluctuation can be revealed only in two-particle quantities, such as the spin susceptibility (Fig.~\ref{fig:W_hf}) or the renormalized interaction (see SM~\cite{SM}). 

Interestingly, similar effects can also be observed away from half-filling, where strong magnetic fluctuations are related to a completely different type of magnetic instability. 
Let us repeat the calculations for the same $t_{2g}$ model~\eqref{eq:Hamiltonian} for the case of ${\av{N_{i}}=4}$ and $U=5$. 
Diagonal and non-diagonal components of the susceptibility are presented in Fig.~\ref{fig:W_dop} and SM~\cite{SM}, respectively. 
For a small value of the Hund's coupling $J=0.2$ (${\lambda=0.71}$) the susceptibility is again nearly diagonal in the orbital space and is highly anisotropic. 
We also find that for the case of $\av{N_{i}}=4$ electrons per lattice site the spatial structure of spin fluctuations is considerably different from the half-filled case and corresponds to an incommensurate spiral state with momentum indicated in Fig.~\ref{fig:W_dop}\,$a$ by the black arrow. 
Nevertheless, the $q_{x}$ direction still remains almost dispersionless. 
Increasing the value of the Hund's coupling to $J=1$, the magnetic l.e. again approaches unity (${\lambda=0.91}$).
Straightforwardly, magnetic fluctuations become isotropic and exhibit a pronounced peak at the center of the Brillouin zone (see Fig.~\ref{fig:W_dop}\,$b$), which is associated with strong ferromagnetic (FM) fluctuations. 
This finding is reminiscent of the case of itinerant FM fluctuations in the $d^{4}$ compounds SrRuO$_3$ in its high-temperature paramagnetic phase~\cite{RevModPhys.84.253, PhysRevB.74.094104}. 

The three-point vertex corrections introduced here are the driving force for the orbitally isotropic state of the spin fluctuations.
In their absence, the self-energy~\eqref{eq:Sigma} and polarization operator~\eqref{eq:Pi} of the $\text{D-TRILEX}$ approach would take the form of a simple ``magnetic'' $GW$+DMFT theory. 
Repeating the same calculations for the half-filled $t_{2g}$ model without vertex corrections, we find that the l.e. of magnetic fluctuations approaches unity (${\lambda=0.99}$) already for a relatively small value of the Hund's coupling ${J=0.4}$ (see SM~\cite{SM}). 
However, the diagonal susceptibility $X^{\rm sp}_{11}$ remains anisotropic in momentum space.
Inserting vertex corrections back, at ${J=0.4}$ the leading eigenvalue reduces to ${\lambda=0.89}$, but the susceptibility $X^{\rm sp}_{11}$ is still anisotropic and looks similar to the one calculated without vertex corrections. 
Finally, increasing the Hund's coupling to $J=0.65$ results in an isotropic form of the diagonal susceptibility $X^{\rm sp}_{11}$, which can be revealed in the regime of strong spin fluctuations only when vertex corrections are considered.
In the case of ${\av{N_{i}}=4}$ the effect of the vertex corrections is even more substantial. 
At large Hund's coupling ${J=1}$ the l.e. of magnetic fluctuations stays nearly the same for both considered approaches (see SM~\cite{SM}).
At the same time, if vertex corrections are neglected the corresponding magnetic mode remains incommensurate as in the regime of small Hund's coupling.
On the contrary, the leading magnetic mode calculated at large ${J=1}$ using the $\text{D-TRILEX}$ approach becomes FM as discussed above.
These result show that the $\text{D-TRILEX}$ approach indeed provides a minimal consistent multi-orbital diagrammatic extension of DMFT that correctly describes collective electronic fluctuations.
In contrast to other methods that rely on more complicated four-point vertex corrections which are numerically expensive and require huge memory size~\cite{PhysRevB.95.115107, doi:10.7566/JPSJ.87.041004, PhysRevB.103.035120}, 
in the $\text{D-TRILEX}$ approach the four-point vertices are eliminated from the theory by using a partially bosonized approximation for the interaction~\cite{PhysRevLett.121.037204, PhysRevB.99.115124, PhysRevB.100.205115}. 
This keeps the method computationally tractable and allows calculations even for multi-orbital systems.

\begin{figure}[t!]
\includegraphics[width=0.95\linewidth]{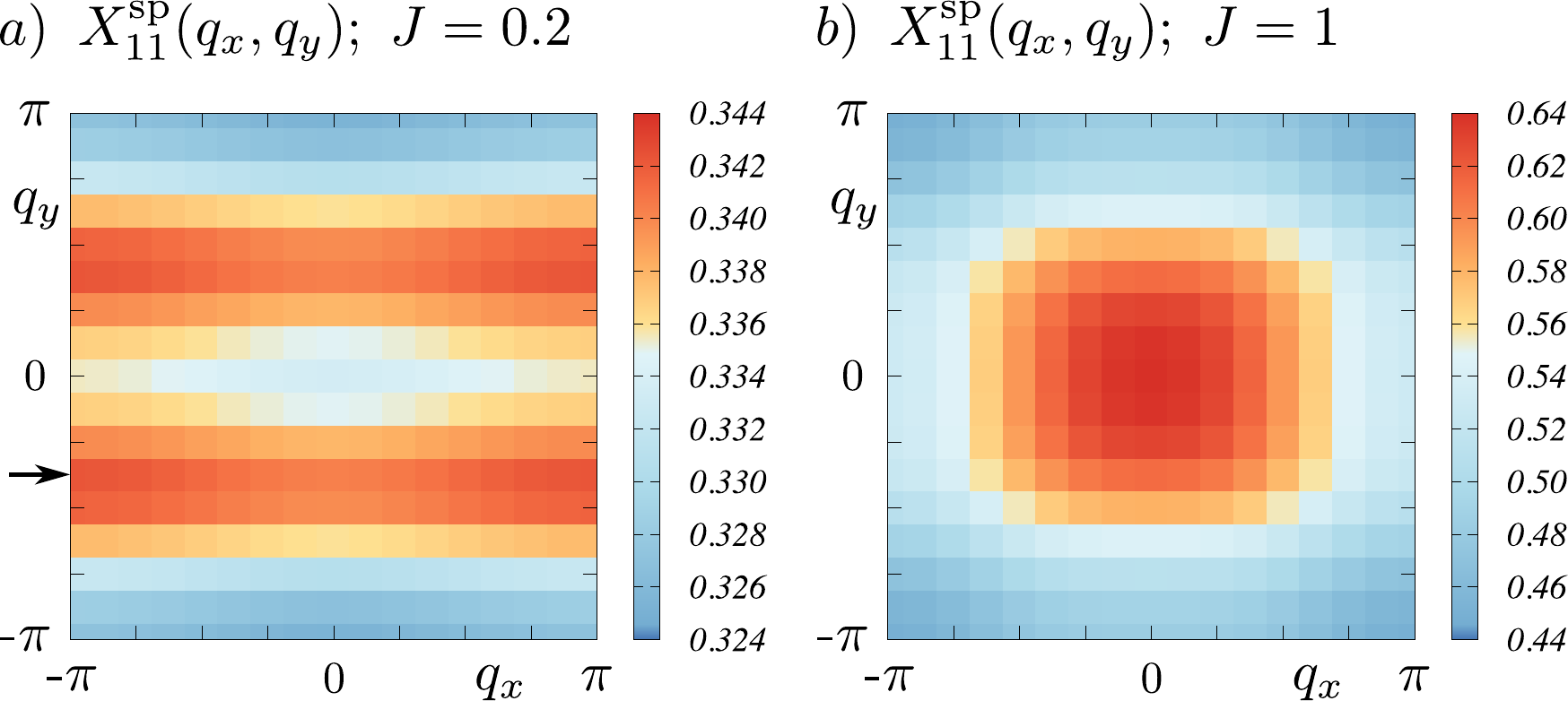}
\caption{\label{fig:W_dop} The absolute value of the diagonal $yz$ orbital component of the spin susceptibility ${X^{\rm sp}_{11}(q_{x}, q_{y}; q_{z}=0, \omega=0)}$ obtained for $U=5$ for the doped $t_{2g}$ model.
(a) For small $J=0.2$, the diagonal susceptibility $X^{\rm sp}_{11}$ is highly anisotropic, and spin fluctuations correspond to an incommensurate spiral state associated with the momentum depicted by the black arrow. (b) For large $J=1$, intraorbital spin fluctuations are isotropic and ferromagnetic as shows the symmetric bright spot at the center of the Brillouin zone.
}
\end{figure}

{\it Conclusions ---}
We have studied collective spin fluctuations in a realistic strongly interacting highly anisotropic three-orbital model. 
We have found that the Hund's coupling enhances collective electronic effects in the spin channel. 
Strong magnetic fluctuations efficiently mix orbital and spatial degrees of freedom leading to orbitally isotropic behavior of the system. 
Remarkably, this effect emerges independently of the antiferromagnetic or ferromagnetic nature of spin fluctuations.
Our findings suggest to revisit the theoretical description of collective excitations in multi-orbital materials with uttermost care.
On the experimental side, neutron scattering experiments should in principle be able to reveal the predicted effects.
The ferromagnetic character of spin fluctuations in SrRuO$_3$ can be interpreted as a special case of this isotropic behavior~\cite{RevModPhys.84.253, PhysRevB.74.094104}.
Calculations incorporating the quasi-2D nature of the electronic structure of Sr$_2$RuO$_4$ provide an interesting perspective for explaining deviations between experimental neutron scattering data and results of current theories~\cite{PhysRevLett.122.047004, PhysRevB.103.104511, Boehnke_2018, PhysRevB.100.125120}.

\begin{acknowledgments}
The authors thank Yvan Sidis, Eva Pavarini, Igor Mazin, and Josef Kaufmann for useful discussions and comments.
The work of E.A.S. is supported by the Russian Science Foundation Grant 18-12-00185.
The work of Y.N. is supported by JSPS KAKENHI 16H06345, 17K14336, 18H01158, and 20K14423.
The work of A.I.L. is supported by European Research Council via Synergy Grant 854843 - FASTCORR, by the Cluster of Excellence ``Advanced Imaging of Matter'' of the Deutsche Forschungsgemeinschaft (DFG) - EXC 2056 - Project No. ID390715994, and by North-German Supercomputing Alliance (HLRN) under the Project No. hhp00042.
S.B. acknowledges support from the French Agence  Nationale  de  la  Recherche  in  the  framework  of the  collaborative  DFG-ANR  project  RE-MAP (Project No.  316912154) and from IDRIS/GENCI Orsay under projet t2020091393.
\end{acknowledgments}

\bibliography{Ref}

\appendix
\clearpage
\onecolumngrid
\begin{center}
\begin{large}
\textbf{
Supplemental Material\\[0.5cm]
Orbital Isotropy of Magnetic Fluctuations in Correlated Electron Materials \\
Induced by Hund's Exchange Coupling}\\[1.5cm]
\end{large}
\end{center}
\onecolumngrid

\section{Nondiagonal component of the spin susceptibility}

\begin{figure}[h!]
\begin{minipage}[t!]{0.48\textwidth}
\includegraphics[width=0.95\linewidth]{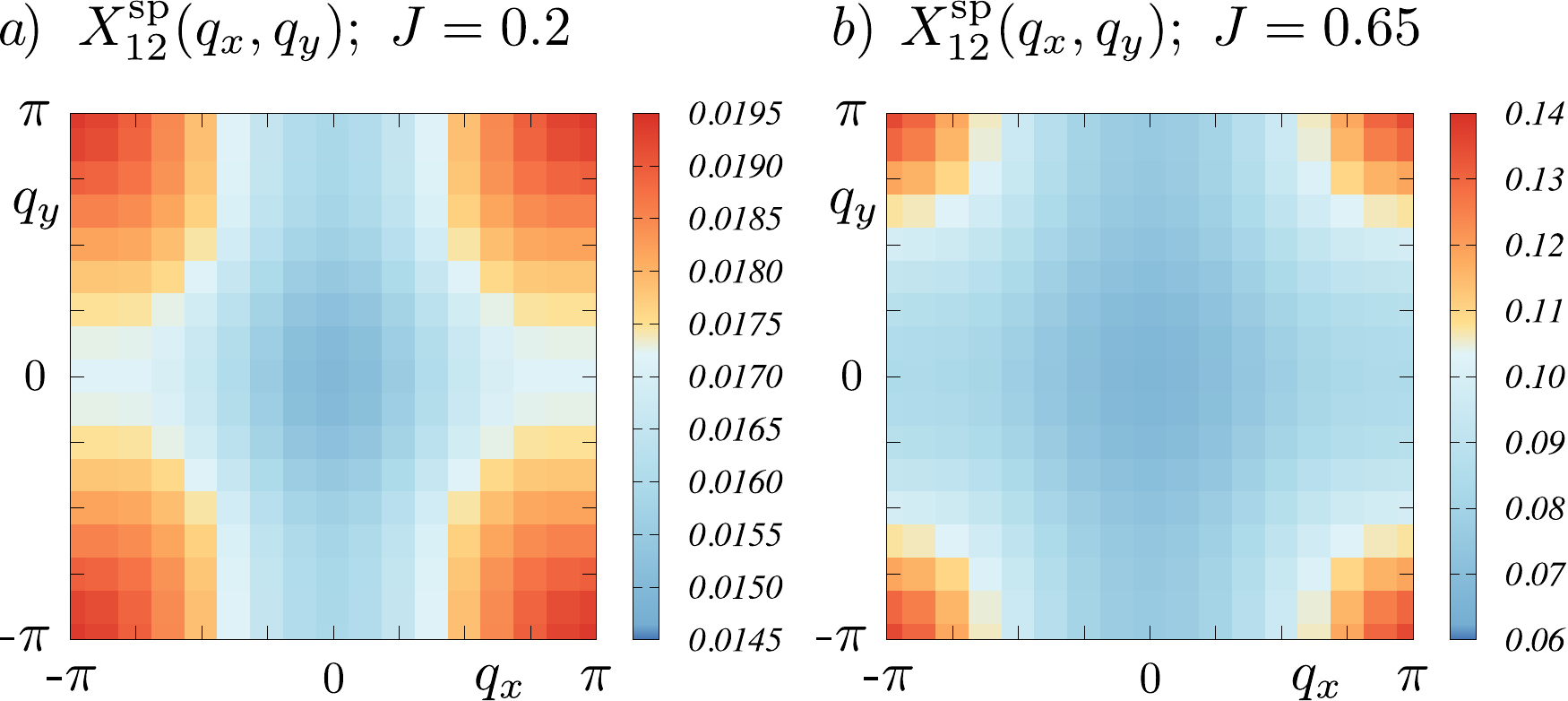}
\caption{\label{fig:X_hf} The absolute value of the nondiagonal component of the spin susceptibility ${X^{\rm sp}_{12}}$ on the ${(q_{x}, q_{y}; q_{z}=0, \omega=0)}$ plane obtained for the half-filled $t_{2g}$ model.}
\end{minipage}
\hspace{0.55cm}
\begin{minipage}[t!]{0.48\textwidth}
\includegraphics[width=0.95\linewidth]{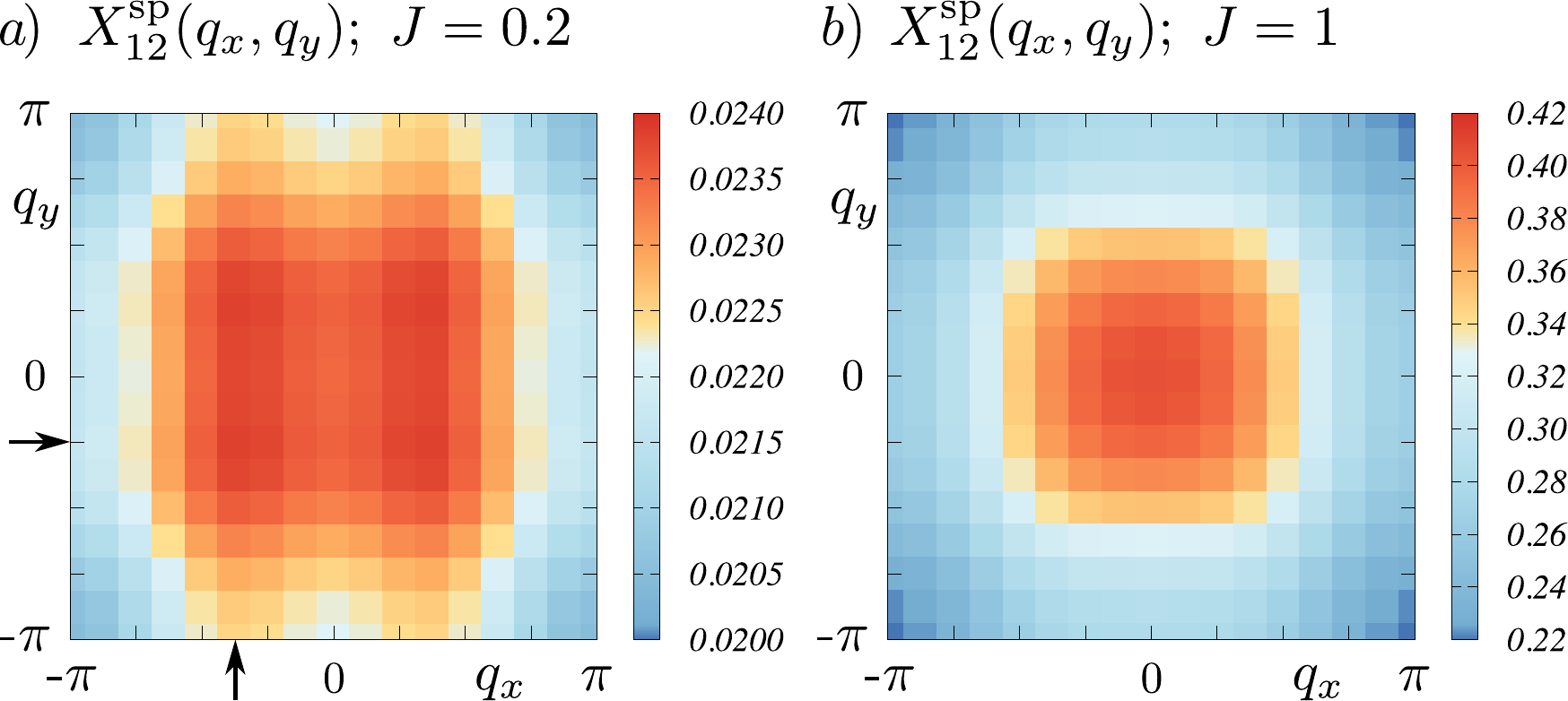}
\caption{\label{fig:X_dop}The absolute value of the nondiagonal component of the spin susceptibility ${X^{\rm sp}_{12}(q_{x}, q_{y}; q_{z}=0, \omega=0)}$ obtained for the case of $4$ electrons per lattice site.}
\end{minipage}
\end{figure}

\vspace{0.25cm}
In this section we show the nondiagonal part of the spin susceptibility $X^{\rm sp}_{12}(q_{x}, q_{y}; q_{z}=0; \omega=0)$, where indices $1$ and $2$ correspond to $yz$ and $zx$ orbitals, respectively. 
In the half-filled case $\av{N_i}=3$ and small Hund's coupling $J=0.2$ (Fig.~\ref{fig:X_hf}\,a) this quantity is almost isotropic, but negligibly small compared to the diagonal component of the susceptibility $X^{\rm sp}_{11}$ shown in the main text. 
On the contrary, for large Hund's coupling $J=0.65$ and the same filling $\av{N_i}=3$ the nondiagonal part of the spin susceptibility is fully isotropic with the pronounced antiferromagnetic behavior (see Fig.~\ref{fig:X_hf}\,b), and its amplitude is comparable to the one of the $X^{\rm sp}_{11}$.
A similar behavior of the nondiagonal spin susceptibility can also be found for the filling of $\av{N_i}=4$ electrons per lattice site shown in Fig.~\ref{fig:X_dop}. In this case, for small ${J=0.2}$ interorbital spin fluctuations correspond to an incommensurate spiral state associated with momenta depicted by black arrows. 
For large ${J=1}$, interorbital spin fluctuations are isotropic and ferromagnetic as shows the symmetric bright spot at the center of the Brillouin zone. 

\section{Nonlocal Green's function at half-filling}

\begin{figure}[b!]
\includegraphics[width=0.45\linewidth]{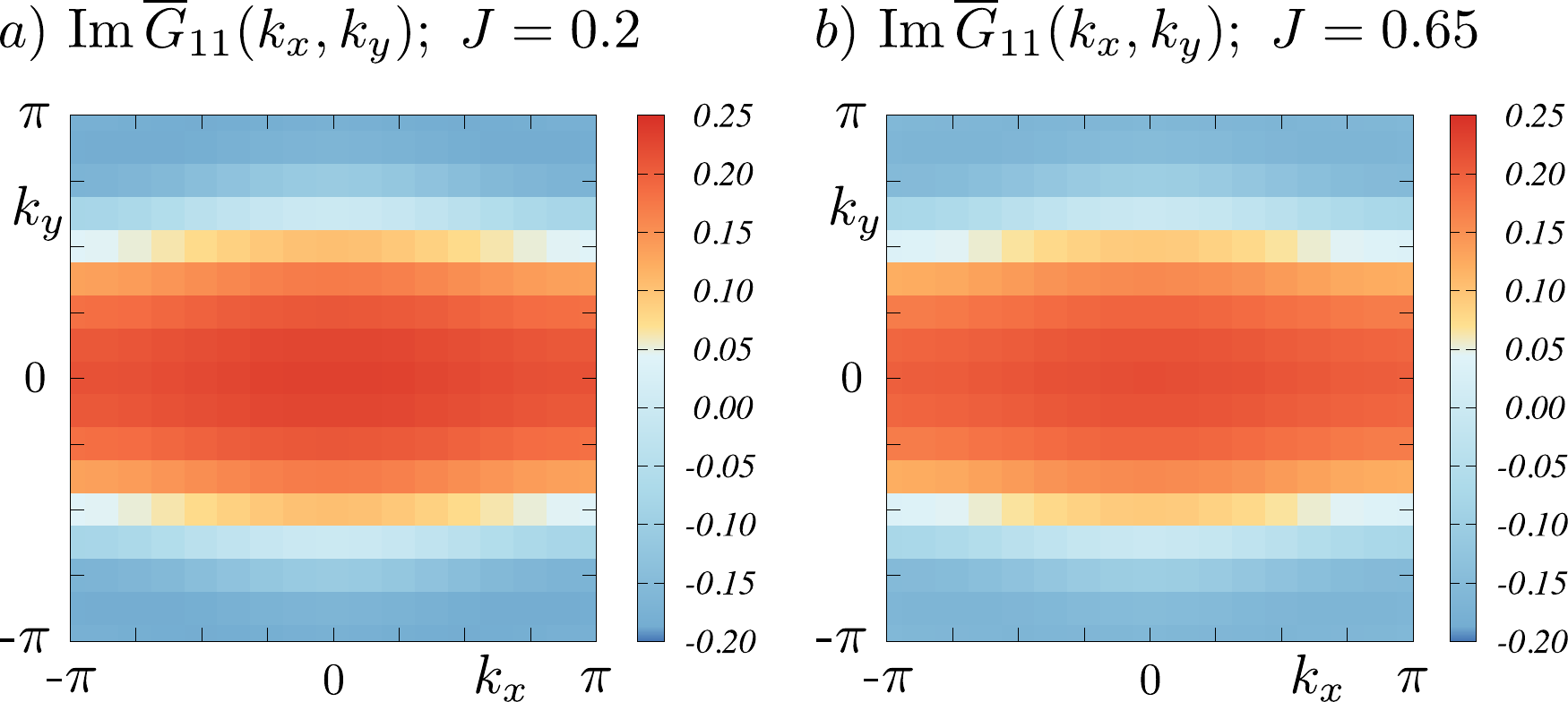}
\caption{\label{fig:GF} Imaginary part of the nonlocal Green's function ${\overline{G}_{11}(k_x, k_y; k_z=0, \nu=\pi/\beta)}$ for the $yz$ orbital obtained for the half-filled $t_{2g}$ model.}
\end{figure}

In this section we show the imaginary part of the nonlocal Green's function obtained for the half-filled case. Fig.~\ref{fig:GF} illustrates that for both values of the Hund's coupling (a) ${J=0.2}$ and (b) ${J=0.65}$ the Green's function remains anisotropic in momentum space, as follows from the discussion presented in the main text of the paper.

\clearpage

\section{Renormalized spin interaction}

\begin{figure}[h!]
\begin{minipage}[t!]{0.48\textwidth}
\includegraphics[width=0.95\linewidth]{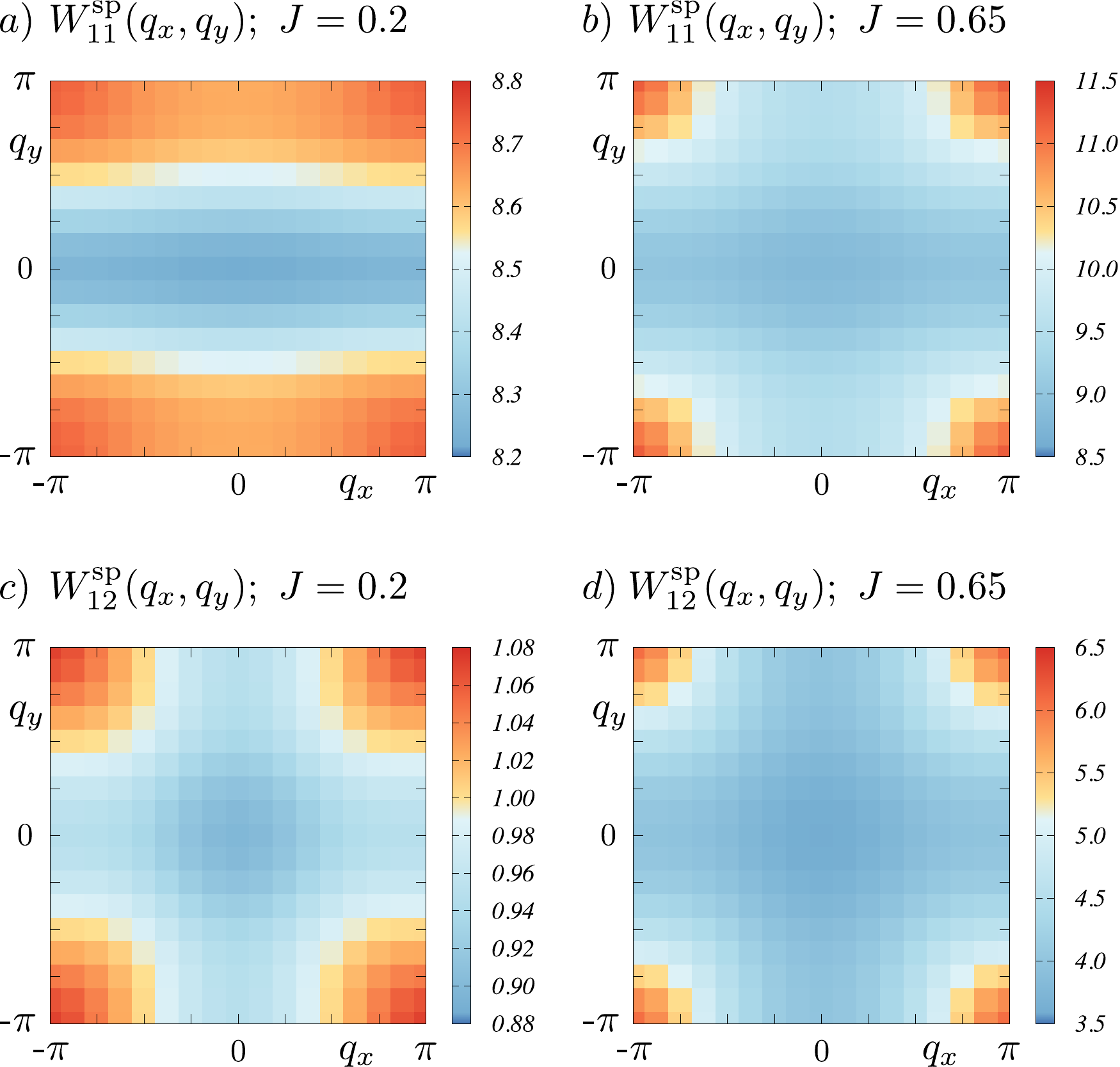}
\caption{\label{fig:W_hf} The absolute value of the diagonal (a, b) and nondiagonal (c, d) components of the renormalized spin interaction ${W^{\rm sp}_{ll'}(q_{x}, q_{y}; q_{z}=0, \omega=0)}$ obtained for the half-filled case.}
\end{minipage}
\hspace{0.55cm}
\begin{minipage}[t!]{0.48\textwidth}
\includegraphics[width=0.95\linewidth]{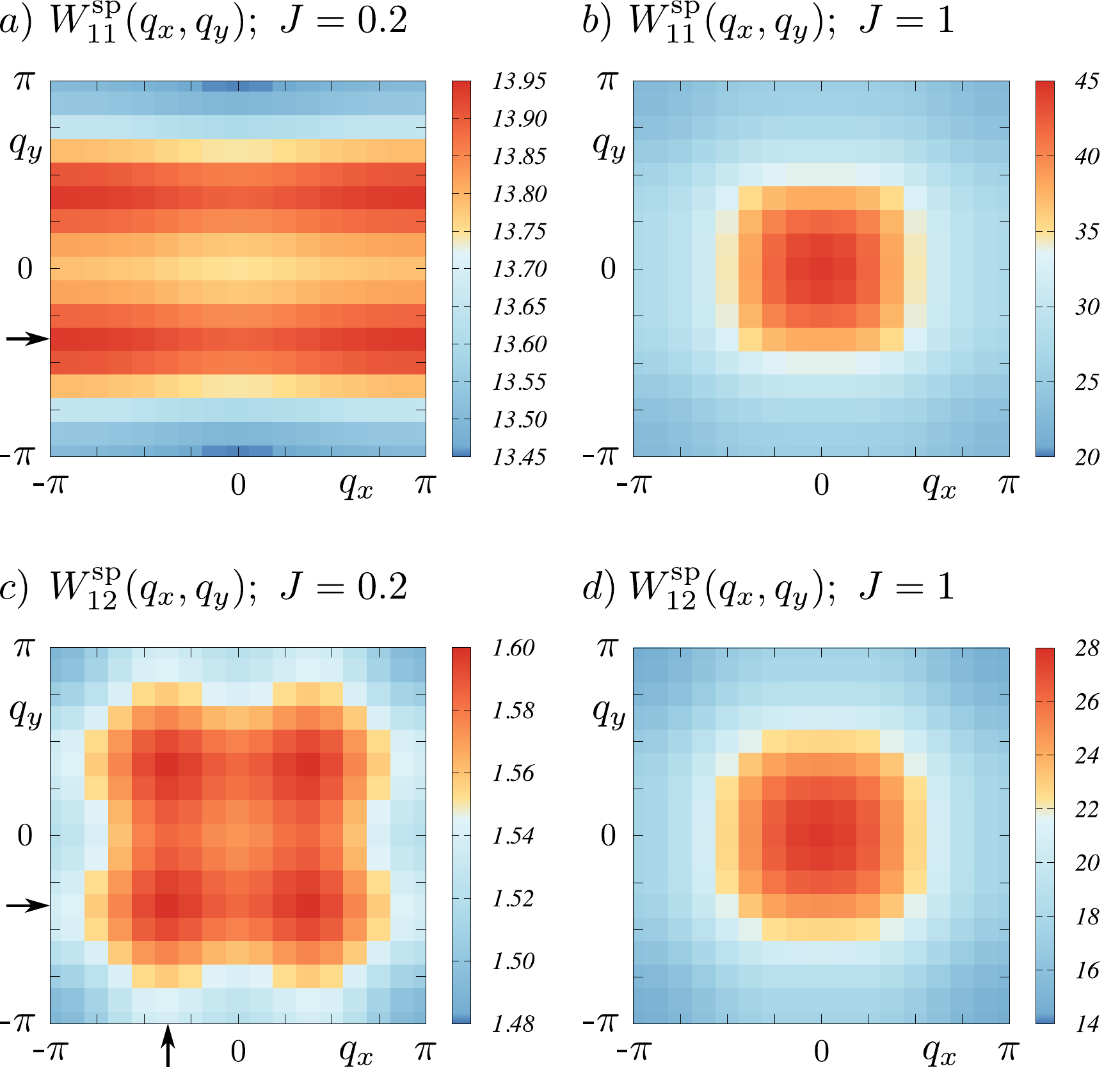}
\caption{\label{fig:W_dop}The absolute value of the diagonal (a, b) and nondiagonal (c, d) components of the renormalized spin interaction ${W^{\rm sp}_{ll'}(q_{x}, q_{y}; q_{z}=0, \omega=0)}$ obtained for the filling $\av{N_i}=4$.
}
\end{minipage}
\end{figure}

\vspace{0.25cm}
The anisotropic to isotropic transition of magnetic fluctuations, which has been captured by the spin susceptibility $X^{\rm sp}_{ll'}$, can also be seen in the renormalized spin interaction $W^{\rm sp}_{ll'}$ that directly enters the electronic self-energy (see Eq.~5 of the main text). Both, $X^{\rm sp}_{ll'}$ and $W^{\rm sp}_{ll'}$ quantities are obtained via similar Dyson equations
\begin{align}
X^{\rm sp\,-1}_{ll'}(q) &= \Pi^{\rm sp\,-1}_{ll'}(q) - U^{\rm sp}_{ll'} \\
W^{\rm sp\,-1}_{ll'}(q) &= U^{\rm sp\,-1}_{ll'} - \Pi^{\rm sp}_{ll'}(q)
\end{align}
For this reason, the structure of the renormalized spin interaction shown in Fig.~\ref{fig:W_hf} for the half-filled case and in Fig.~\ref{fig:W_dop} for the case of $\av{N_i}=4$ electrons per lattice site is very similar to the one of the spin susceptibility.
Moreover, one may notice that the nondiagonal part of the renormalized spin interaction $W^{\rm sp}_{12}$ for both values of the Hund's coupling, as well as the diagonal interaction $W^{\rm sp}_{11}$ at large $J$ look more isotropic compared to the corresponding components of the spin susceptibility. This observation can be explained by the fact that the local bare interaction $U^{\rm sp}_{ll'}$ gives the first-order contribution to the renormalized interaction $W^{\rm sp}_{ll'}(q)$.
On the contrary, the first-order contribution to the spin susceptibility $X^{\rm sp}_{ll'}(q)$ is given by the polarization operator $\Pi^{\rm sp}_{ll'}(q)$, which in this particular model is anisotropic as discussed in the main text of the paper.

\begin{figure}[b!]
\includegraphics[width=0.7\linewidth]{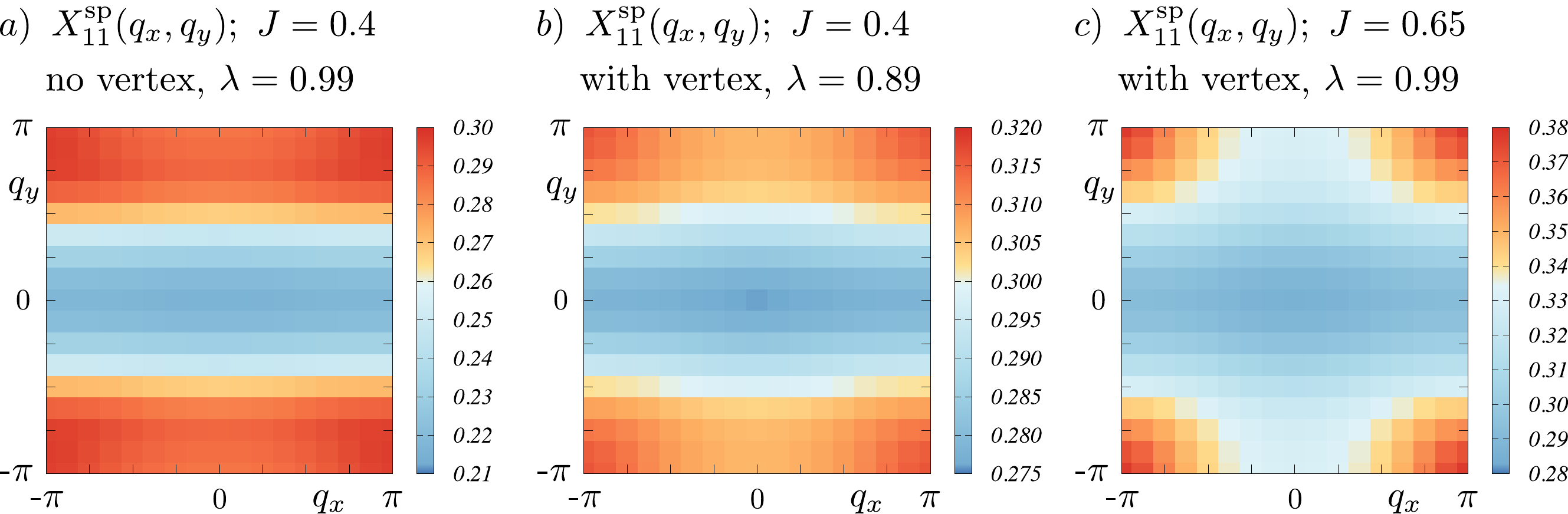}
\caption{\label{fig:X11_U4_SM} 
The absolute value of the diagonal component of the spin susceptibility ${X^{\rm sp}_{11}(q_{x}, q_{y}; q_{z}=0, \omega=0)}$ calculated for the half-filled case.   
}
\end{figure}

\section{Effect of vertex corrections}

In this section we illustrate the importance of vertex corrections considered in the $\text{D-TRILEX}$ diagrams for the self-energy (Eq.~5 in the main text) and polarization operator (Eq.~6 in the main text).
To this aim we first obtain the spin susceptibility for the half-filled case without (Fig.~\ref{fig:X11_U4_SM}\,a) and with (Fig.~\ref{fig:X11_U4_SM}\,b,\,c) vertex corrections. 
Without vertices, the leading eigenvalue of magnetic fluctuations approaches unity (${\lambda=0.99}$) already for a relatively small value of the Hund's coupling ${J=0.4}$.
As Fig.~\ref{fig:X11_U4_SM}\,a shows, in this case the diagonal part of the spin susceptibility $X^{\rm sp}_{11}$ still remains anisotropic in the ${(q_{x}, q_{y}; q_{z}=0)}$ plane. 
Including vertex corrections, the leading eigenvalue reduces to ${\lambda=0.89}$ for the same value of the Hund's coupling ${J=0.4}$. 
The corresponding spin susceptibility $X^{\rm sp}_{11}$ shown in Fig.~\ref{fig:X11_U4_SM}\,b also remains anisotropic and looks similar to the one calculated without vertex corrections. 
Increasing the Hund's coupling to ${J=0.65}$ allows one to obtain the isotropic form of the spin susceptibility $X^{\rm sp}_{11}$, which can be revealed in the regime of strong spin fluctuations (${\lambda=0.99}$) {\it only} if vertex corrections are considered.

\begin{figure}[t!]
\includegraphics[width=0.45\linewidth]{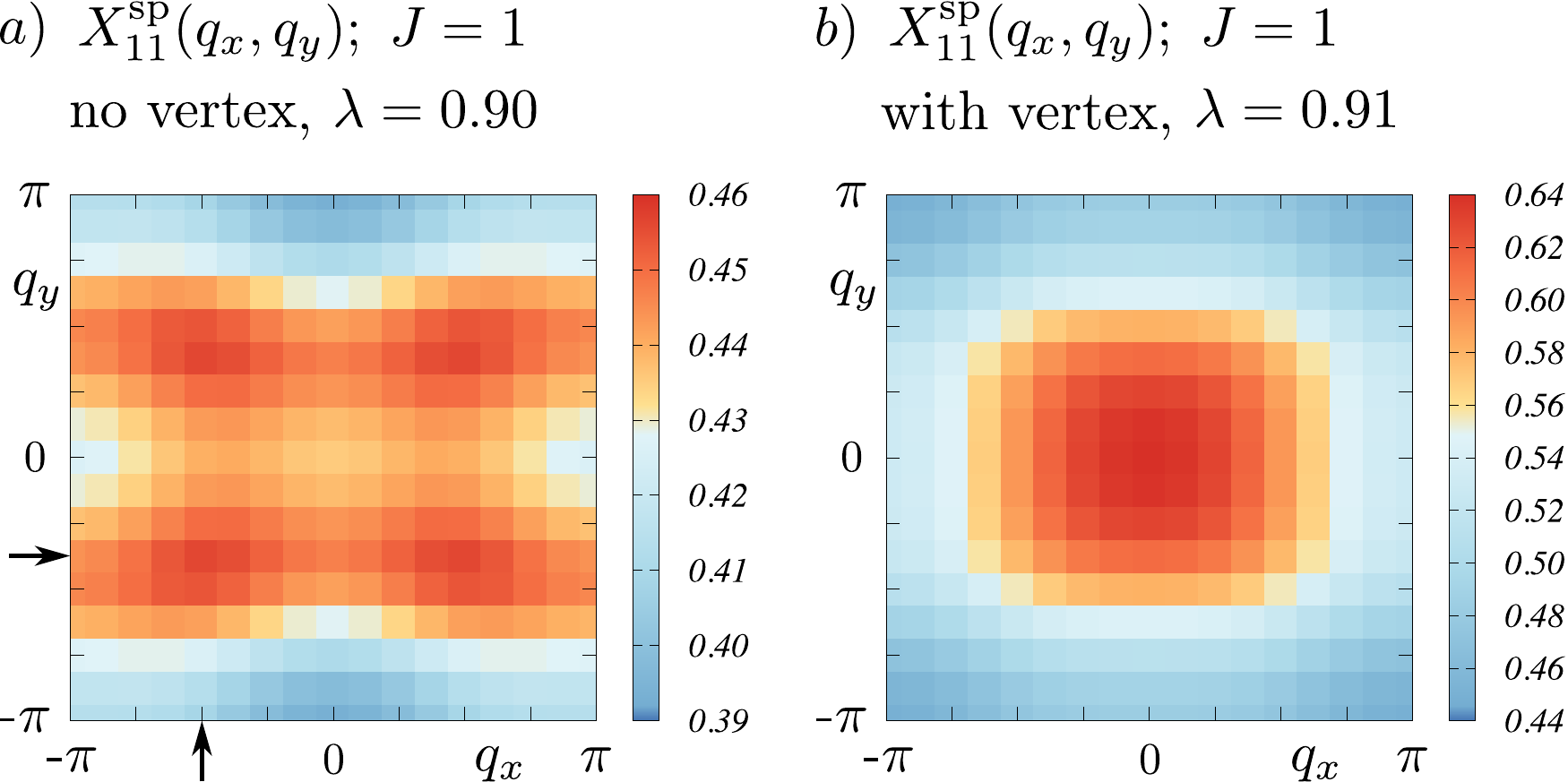}
\caption{\label{fig:X11dopU5_SM} 
The absolute value of the diagonal component of the spin susceptibility ${X^{\rm sp}_{11}(q_{x}, q_{y}; q_{z}=0, \omega=0)}$ calculated for the case of $\av{N_i}=4$ electrons per lattice site and ${J=1}$.  
}
\end{figure}

In the case of ${\av{N_{i}}=4}$ electrons per lattice site the effect of the vertex corrections is even more demonstrative.  
Calculating the spin susceptibility for large Hund's coupling $J=1$ without (Fig.~\ref{fig:X11dopU5_SM}\,a) and with (Fig.~\ref{fig:X11dopU5_SM}\,b) vertex corrections in diagrams for the self-energy and polarization operator one finds approximately the same leading eigenvalue of magnetic fluctuations. 
At the same time, without vertices spin fluctuations correspond to an incommensurate spiral state associated with the momentum depicted by the black arrows in Fig.~\ref{fig:X11dopU5_SM}\,a. 
Inclusion of vertex corrections drastically changes the symmetry of the spin fluctuations, and the leading magnetic mode becomes ferromagnetic  (Fig.~\ref{fig:X11dopU5_SM}\,b). 

\end{document}